# Nanovortex-driven all-dielectric optical diffusion boosting and sorting concept for lab-on-a-chip platforms

**Adrià Canós Valero[1], Denis Kislov[1], Egor A. Gurvitz[1], Hadi K. Shamkhi[1], Dmitrii Redka[2], Sergey Yankin[3], Pavel Zemánek[4] and Alexander S. Shalin[1]**

[1]ITMO University, Kronverksky prospect 49, 197101, St. Petersburg, Russia

[2]Electrotechnical University "LETI" (ETU) 5 Prof. Popova Street, 197376, Saint Petersburg, Russia

[3] LLC COMSOL, Bolshaya Sadovaya St. 10, 123001, Moscow, Russia

[4]Czech Academy of Sciences, Institute of Scientific Instruments, Královopolská 147, 612 64 Brno, Czech Republic

Corresponding author: adria.canos@optomech.ifmo.ru

**Abstract.** The ever-growing field of microfluidics requires precise and flexible control over fluid flow at the micro- and nanoscales. Current constraints demand a variety of controllable components for performing different operations inside closed microchambers and microreactors. In this context, novel nanophotonic approaches can significantly enhance existing capabilities and provide new functionalities via finely tuned light-matter interaction mechanisms. Here we propose a novel design, featuring a dual functionality on-chip: boosted optically-driven particle diffusion and nanoparticle sorting. Our methodology is based on a specially designed high-index dielectric nanoantenna, which strongly enhances spin-orbit angular momentum transfer from an incident laser beam to the scattered field. As a result, exceptionally compact, subwavelength optical nanovortices are formed and drive spiral motion of peculiar plasmonic nanoparticles via the efficient interplay between curled spin optical forces and radiation pressure. The nanovortex size is an order of magnitude smaller than that provided by conventional beam-based approaches. The nanoparticles mediate nano-confined fluid motion enabling nanomixing without a need of moving bulk elements inside a microchamber. Moreover, precise sorting of gold nanoparticles, demanded for on-chip separation and filtering, can be achieved by exploiting the non-trivial dependence of the curled optical forces on the nanoobjects' size. Altogether, this study introduces a versatile platform for further miniaturization of moving-part-free, optically driven microfluidic chips for fast chemical synthesis and analysis, preparation of emulsions, or generation of chemical gradients with light-controlled navigation of nanoparticles, viruses or biomolecules.

## 1. Introduction

Micro-optofluidics represents one of the most promising and fast growing directions in current state-of-the-art science and engineering [1–4]. In particular, the control of fluid flows in microsized channels plays an essential role for applications ranging from the transport of reduced amounts of hazardous or costly substances and DNA biochip technology, to miniaturized analytical and synthetic chemistry [5–9].

The multidisciplinary nature of microfluidics has brought together seemingly unrelated fields, such as electrical and mechanical engineering, biology, chemistry and optics. For example, in the context of chemical engineering, the utilization of distributed microreactors working in parallel can enhance production significantly and facilitates the design of new products [10,11]. However, slow mixing processes constitute a bottleneck that restricts reaction processes, especially when the desired reaction rate is high [12,13]. For this purpose, fast mixing is highly required to avoid the reactive process being delayed by this critical step, and to reduce potential side products [13].

Given the low Reynolds numbers at which fluid flow occurs in microreactors, fluid mixing represents a significant challenge [14–16]. In the most conventional situation where only passive mixing happens, the main driving mechanism corresponds to diffusion (Brownian motion) [14] implying mixing to take place at a very low rate. Consequently, the effective distance that the molecules of a fluid need to travel in a mixer before interacting with another fluid with different composition (i.e. - the mixing length) becomes restrictively long [16]. Passive mixers depend solely on decreasing the mixing length by optimizing the flow channel geometry in order to facilitate diffusion [14,17]. In contrast, active schemes rely on external sources injecting energy into the flow in order to accelerate mixing and diffusion processes and drastically decrease the mixing lengths [15,18].

Most early studies related to micromixers have been focused on the passive type. Conversely, despite their higher cost and complex fabrication methods, the enhanced efficiency of active micromixers with respect to passive ones has drawn the attention of the scientific community in the recent years [15]. Because of the power and size constraints involved in microfluidics, research efforts have been focused on the utilization of mixing principles not involving moving mechanical parts such as surface tension-driven flows [19], ultrasound and acoustically induced vibrations [20,21], and electro- and magneto-hydrodynamic action [16,22].

Given the small operation scales of microfluidics, micron-scale focusing of laser beams, as well as different types of light matter interactions make possible to provide sufficiently strong optical forces to propel particles [23–25], sort objects according to their size or optical properties [26–28] or self-arrange colloidal particles into optically bound structures [23–25,29,30]. Nowadays the additional degrees of freedom offered by complex shaping of laser beams [31–35] have made possible the manipulation and trapping of large amounts of microparticles[36]. In particular, they allow to create optical vortices with helical phase front (e.g. Laguerre-Gaussian or higher-order Bessel-beams) carrying both linear and angular momentum[37,38]. When such an optical micro-vortex is scattered by particles, it induces an optical torque on them leading to their orbital motion around the focus of the laser beam[39]. Due to the angular momentum conservation, elastic scattering of a circularly polarized beam possessing spin-angular momentum[40], by optically anisotropic [41–43] or non-spherical objects[44,45] leads to their spinning oriented along the direction of propagation of the incident illumination. Combining both types of optical angular momentum leads to complex spin-orbital interaction [38,46] and novel interesting phenomenon, e.g. detection of spin forces[47,48].

At the nanoscale, metal-based plasmonics dominates and provides exciting means for trapping and manipulating nanoobjects [49–52]. On the other hand, the recently growing field of all-dielectric nanophotonics[53] presents itself as a promising alternative for the integration of optomechanical concepts in microfluidic devices. Properly designed dielectric nanostructures with finely tuned Mie resonant response provide the means for tailoring electric and magnetic components of the scattered light [53–55]. They allow to obtain strong near fields, which induce substantial optical forces acting upon other subwavelength scatterers dispersed in the medium surrounding the nanostructure [56–60].

In this work, we focus on the conversion of spin angular momentum (SAM) of an incident circularly polarized plane wave into orbital angular momentum (OAM[61,62]) of the scattered field mediated by a specially designed nanostructure constituted of a realistic high refractive index material (silicon). In contrast to the above-mentioned methods, the optical vortex field created in this way is very localized, only reaching a few hundreds of nanometers in diameter. Moreover, the induced optical forces are strong enough to propel gold (Au) nanoparticles of particular sizes along spiral trajectories around the nanostructure. Based on the latter effect, we propose a novel method for mixing fluid in nanovolumes mediated by chemically inert Au nanoparticles (see Figure 1a). In addition, we take advantage of the size sensitivity of the Au polarizability to achieve light-mediated nanoparticle separation by means of the same geometrical configuration (see Figure 1b). The subwavelength size of the investigated optical nanovortex greatly enhances the length scale of interaction in comparison to the more conventional approaches involving Bessel beams[63,64], opening new directions in light-matter interaction via light angular momentum exchange. We believe that the proposed simple geometry for optically driven diffusion boosting and nanoparticle sorting is of high interest for a plethora of applications in microfluidics and lab-on-a-chip devices.

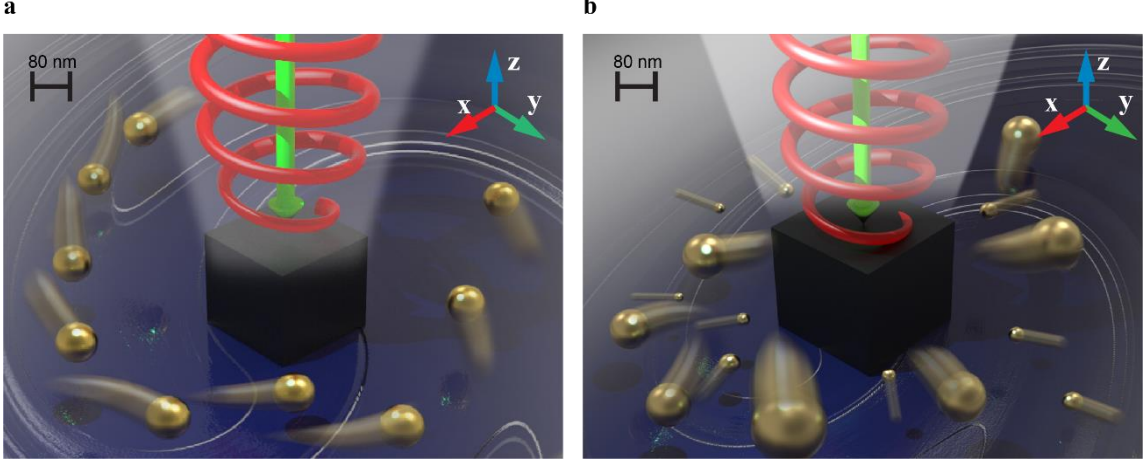

**Figure 1.** An artistic view of the proposed active nanomixing scheme (left) and radial separation of nanoparticles (right). (a) A silicon nanocube submerged in a water solution is illuminated by a circularly polarized laser beam coming from the top. The scattered field carries a nonzero tangential component of the pointing vector in the xy plane, which induces nonzero orbital angular momentum in the negative z direction. The same effect causes the spiral motion of Au nanoparticles around the nanocube. Viscous friction between the moving nanoparticles and the fluid gives rise to convective fluid motion and enhances fluid mixing. (b) Sorting concept. Nanoparticles of different sizes having opposite signs of the real part of polarizability are radially displaced in opposite directions – the smaller ones move towards the nanocube while larger ones move away from it.

## 2. Formation of an optical nanovortex

The spiral motion of nanoobjects in an optical nanovortex driven by an out-of-plane light source (Fig. 1a) requires, on the one hand, efficient transformation of spin angular momentum (SAM) of light to in-plane orbital angular momentum (OAM) of the highly confined scattered near fields of the nanocube, which, in turn, should be transferred to the nanoparticles. Therefore, as a first constraint, sufficient in-plane scattering from the nanocube should take place. This urgent functionality could be enabled, in particular, by the recently observed Transverse Kerker Effect [65] allowing for lateral-scattering only. On the other hand, azimuthal forces arising due to helicity inhomogeneities in the scattered near field (curl spin forces[66]) also enable rotational motion. Hereinafter, we optimize both effects taking into account that we, actually, do not require the total suppression of forward and backward scattering as in [65], and, therefore, we can tune the parameters in order to obtain an enhanced optical subwavelength vortex.

A cubical Si nanostructure with refractive index $n \approx 4$ (e.g., silicon at the visible range) and edge length 250 nm is illuminated by a circularly polarized plane wave propagating along the negative $z$-axis (see inset of Figure 2a). For such an incident field one can calculate the angular momentum flux density using the expressions for paraxial waves [67]. The $z$ component can then be written in the general form

$$J_z(\mathbf{r}) = \frac{c\varepsilon_0}{2\omega}\left(\mathbf{E}_{inc}^* \cdot \mathbf{L}\mathbf{E}_{inc}\right)\bigg|_z + 2c\langle \mathbf{L}_s \rangle\big|_z , \qquad (1)$$

where $\mathbf{E}_{inc}$ is the electric field, $\varepsilon_0$ is the vacuum permittivity, $\mathbf{L}$ is the orbital angular momentum operator[68], and $\langle \mathbf{L}_s \rangle = \varepsilon_0 n_m^2 \mathbf{E} \times \mathbf{E}^* / (4i\omega)$ is the electric contribution to the SAM flux density[69] of the beam in a medium with refractive index $n_m$. The first term in the right-hand side of (1) corresponds to the OAM carried by $\mathbf{E}_{inc}$. Substituting the expression of a circularly polarized plane wave into (1) yields the OAM term equal to zero and the total angular momentum flux density is entirely given by the SAM flux density:

$$J_z = \sigma \frac{I_0}{\omega}, \qquad (2)$$

where the wave helicity $\sigma$ takes the values of +1 and -1 for left and right-circular polarization, respectively, and $I_0$ is the incident light intensity. Since the nanostructure has negligible losses, the total angular momentum of incident and scattered light is conserved. This conservation law for the total angular momentum implies that the part of the incident SAM, given by Eq. (2), is transferred to both SAM and OAM of the scattered field.

We can write the total angular momentum surface density of the scattered wave in full analogy with classical mechanics [68] as

$$\langle \mathbf{J} \rangle = \frac{\mathbf{r} \times \langle \mathbf{S}^s \rangle}{c}, \qquad (3)$$

where $\langle \mathbf{S}^s \rangle$ denotes the time-averaged scattered Poynting vector. Since $J_z$ is non-zero due to Eq.(2), Eq. (3) implies that the Poynting vector of the scattered field has non-zero tangential components.

The transverse components of the Poynting vector in the field scattered by the nanocube display similar rotating features as for chiral scatterers such as helices or gammadion-like structures [70–72], however the fabrication process of an isotropic nanocube is much less complex and no chirality is involved. In order to gain better physical insight, we consider the multipole decomposition for the scattering cross-section of the nanoparticle illuminated with LCP light deposited over a glass substrate (Figure 2a) [73], and find the most pronounced vortex-like energy flux at a close vicinity to transverse Kerker point ($\lambda = 788$ nm) – at the magnetic quadrupole (MQ) resonance (Figure 2b, $\lambda = 765$ nm). Figure 2a presents the result in the form of the spectral dependence of the scattering cross-sections of individual multipoles and the total cross-sections.

In particular, the magnetic quadrupole (MQ) mode presents a very high signal-to-noise-ratio with respect to the other leading multipoles; the magnetic dipole and electric quadrupole are almost one order of magnitude smaller and are both out of resonance, while the electric dipole radiation is suppressed by an anapole state coinciding with the MQ resonant frequency. Thus, "pure" MQ fields can be obtained, which provide stronger near-field effects in comparison with lower quality resonances. The resonant MQ mode displaying negligible contributions of the other multipoles enhances the vorticity of the Poynting vector[74] (Figure 2b).

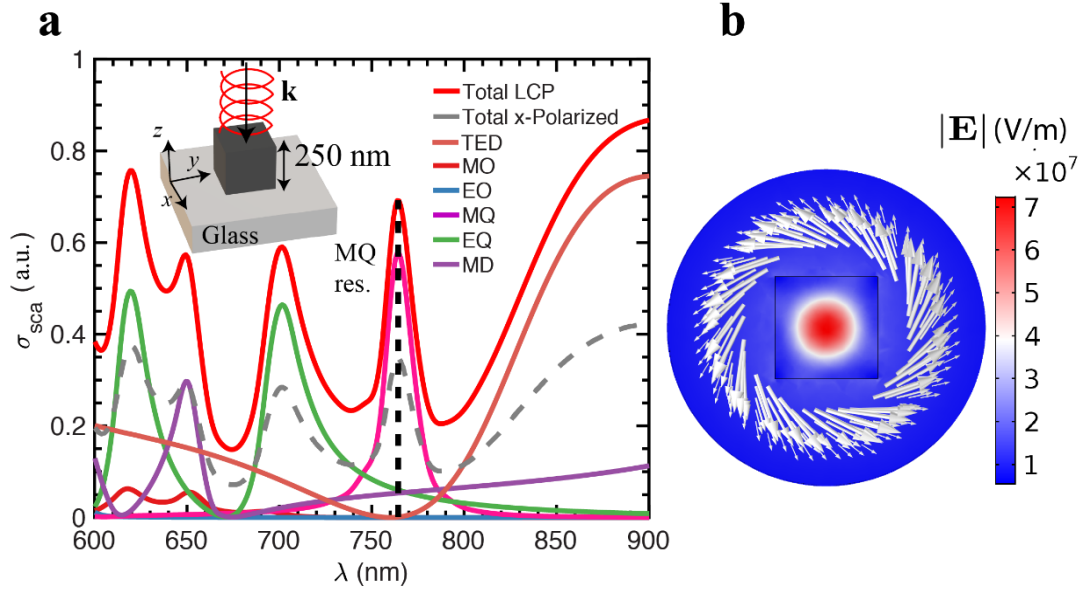

**Figure 2.** (a): Cartesian multipole decomposition of the scattering cross-section of the dielectric cube deposited on a glass substrate and centered at the origin of the coordinate system; the cube is illuminated by a left-hand circularly polarized plane wave propagating against the z-axis. The geometry is illustrated in the top inset and the ambient medium is air. The dashed black line indicates the position of the resonant MQ mode. The dashed grey curve corresponds to the total scattering cross section for linearly polarized incident wave. The total scattered power with incident LCP illumination is obtained from the sum of contributions of individual multipoles (Total LCP). (b) Colorplot denotes the norm of the total electric field at the resonant MQ wavelength in the transverse x-y plane at z=100 nm. The arrows indicate direction and their lengths illustrate the relative size of the transverse part of the Poynting vector.

While the numerical results shown in Figure 2 provide a clear link between the enhanced transfer of incident field SAM to scattered field OAM at the MQ resonance, a complete physical picture requires a deeper theoretical insight on the behavior of the fields produced by the MQ mode under the prescribed illumination. For that purpose, we now express the amount of power extracted from the field by the nanocube (referred to as the extinction power) [68] as

$$P_{ext} = \frac{1}{2}\text{Re}\left\{\int_{V_p} \mathbf{E}_{inc}^* \cdot \mathbf{j}(\mathbf{r}) d^3\mathbf{r}\right\}, \quad (4)$$

where the integration is carried inside the volume of the nanocube $V_p$ of the nanocube, and $\mathbf{j}(\mathbf{r})$ is the induced current. We consider $\mathbf{E}_{inc}$ to be a left-hand circularly polarized plane wave with the form $\mathbf{E}_{inc}(\mathbf{r}) = \tilde{E}_0(z)(\mathbf{e}_x + i\mathbf{e}_y)$, with $\tilde{E}_0(z) = E_0 e^{-ik_0 z}$. Performing the multipole expansion of $\mathbf{j}(\mathbf{r})$ in Eq. (4)[75] in the Cartesian basis and substituting the expression of $\mathbf{E}_{inc}$, we obtain the extinction power of the MQ mode

$$P_{ext} = \frac{k_0^2}{4}\text{Re}\left\{M_{zx}E_y^*(0) - M_{zy}E_x^*(0)\right\}, \quad (5)$$

where $M_{ij}$ is $ij$-th component of the magnetic quadrupole tensor and $E_x, E_y$ indicate the components of $\mathbf{E}_{inc}$, respectively. The center of the nanocube is placed at the origin of the coordinate system, with the $x$, $y$ and $z$-axis oriented perpendicular to its sides. Due to its inherent rotational symmetry, the optical response of the nanocube is identical for plane waves linearly polarized in the $x$ or $y$-axis, which gives $|M_{zy}| = |M_{zx}|$, in full agreement with the numerical results presented in Figure 2a. Therefore, under this condition, and neglecting reflection from the glass substrate, Eq. (5) can be simplified as

$$P_{ext} = -\frac{E_0 k_0^2}{2}\text{Im}\{M_{zx}\}. \quad (6)$$

Several conclusions can be readily drawn from Eqs. (5) and (6). Firstly, due to the chosen geometry, only the $M_{zx}$ and $M_{zy}$ components of the magnetic quadrupole tensor are excited (since we work in the irreducible Cartesian multipole representation, the multipole tensors are symmetric and therefore the $M_{xz}$ and $M_{yz}$ components are also excited). Secondly, due to the symmetry of the system, the contributions of $M_{zx}$ and $M_{zy}$ components to the extinction over an oscillation period are equal. Therefore, the total extinction power at the MQ resonance excited by the circularly polarized incident wave is exactly two times larger comparing to a linearly polarized incident wave. Since we assume negligible absorption, the conclusions made for extinction are also valid for scattering cross sections shown in Figure 2a. The intuitive physical picture is the following: during an oscillation period, the incident circularly polarized electric field gradually changes its polarization between the x and y-axis. Consequently, the components of the excited magnetic quadrupole tensor oscillate accordingly. The scattered electric field receives contributions from two magnetic quadrupole moments $\pi/2$ delayed from each other. In analogy with a rotating electric dipole [76], the scattered near-field at the MQ resonance can be obtained as a superposition of the fields generated by $M_{zy}$ and a $-\pi/2$ delayed $M_{zx}$ component with equal amplitudes. Due to the phase delay, the scattered Poynting vector acquires a nonzero tangential component $S_\phi^s$ [77]:

$$S_\phi^s = \sigma \frac{3|M_{zx}|^2}{16\pi^2 c\mu_0} \frac{(9+3r^2k_0^2+r^4k_0^4)}{k_0 r^7} \cos(\theta)^2 \sin(\theta), \tag{7}$$

where $\theta$ is the polar angle in spherical coordinates, and $r=|\mathbf{r}|$. In the x-y plane $\theta = \pi/2$ and $S_\phi^s = 0$. The total Poynting vector, however, also includes an interference term between the scattered electromagnetic field and the incident one, which leads to non-negligible curl in the x-y plane. This is indeed what is observed in the numerical simulations (see Figure 3a, where $\Gamma_z^P$ is proportional to $\langle S_\phi \rangle$). Substituting the scattered Poynting vector in Eq. (3), the time-averaged scattered angular momentum density component in the z-axis $J_z$ can be determined as

$$J_z = \sigma \left| \frac{r}{c} S_\phi^s \right|. \tag{8}$$

Equation (8) provides direct evidence that SAM from the incident wave has been transferred to the scattered field giving rise to the optical vortex shown in Figure 2b. Moreover, since $J_z$ depends on the choice of origin of the coordinate system [78] it can be directly correlated with the extrinsic OAM of the scattered field. Further inspection of Eqs. (7) and (8) also shows that the tangential component of the Poynting vector as well as the angular momentum scale quadratically with the amplitude of the MQ moment, enhancing the field vorticity at the MQ resonance. Since the angular momentum scales as $r^{-n}$ (where $n$ is a positive integer) in the near field, the vorticity of the Poynting vector is very high close to the particle, but decreases very fast going away from it, as confirmed in Figure 2b. The latter has very important consequences regarding the optical forces governing the motion of plasmonic nanoparticles under the influence of such a field, as we study below.

In this section, we have provided analytical expressions describing the excitation of the MQ mode with circularly polarized incident light and proposed an intuitive physical picture explaining the multipolar origin of the tangential component of the Poynting vector giving rise to an optical vortex in the near field of the nanocube.

**3. Optical nanovortex-mediated forces and torques**

We can now proceed to study the effect of the scattered field on small dipolar particles. The time-averaged optical force $\langle \mathbf{F}_o \rangle$ acting upon an induced electric dipole (a nanoparticle), illuminated with the optical nanovortex field can be written as [69]

$$\langle \mathbf{F}_o \rangle = \frac{\alpha'}{2} Re\{\nabla \mathbf{E} \cdot \mathbf{E}^*\} + \frac{ck_0 n_m \alpha''}{\varepsilon_0}\left(\frac{1}{c^2}\langle \mathbf{S} \rangle + \nabla \times \langle \mathbf{L}_s \rangle\right), \quad (9)$$

where $n_m$ is the refractive index of the host medium, $\mathbf{E}$ is the sum of the incident and scattered electric fields, and $\alpha'$ and $\alpha''$ are the real and imaginary parts of the particle dipole polarizability, respectively. The first term on the right-hand side of Eq. (9) corresponds to conservative (curl-free) gradient optical forces acting upon the nanoparticle, which for positive $\alpha'$ drags the nanoparticle towards the region of maximal field intensity. The terms in round brackets describe non-conservative or "scattering" optical forces, hereinafter noted as $\langle \mathbf{F}_{sc} \rangle$. The latter receives contributions from the total Poynting vector $\langle \mathbf{S} \rangle$ and the electric field contribution to the SAM flux density $\langle \mathbf{L}_s \rangle$ [69]. A MQ mode corresponds to an object of well-defined parity, i.e. a transverse electric (TE) multipole. At the resonance, pure electric or magnetic multipoles strongly break electromagnetic duality, and, consequently, do not present a well-defined helicity[74,79]. This effect manifests itself strongly in the near field[66], and implies that the SAM flux density, which is linked to the helicity density[74], features a non-uniform spatial distribution. Therefore, the second term in $\langle \mathbf{F}_{sc} \rangle$, acknowledged as the curl spin force[66], is not only non-negligible but also contributes significantly to the total force and torque exerted on dipolar particles. Employing Eqs. (3) and (9), the $z$ component of the optical torque $\Gamma_z$ acting upon the nanoparticle due to $\langle \mathbf{F}_{sc} \rangle$ is found to be proportional to the tangential components of the Poynting vector $\langle S_\phi \rangle$ and the curl of $\langle \mathbf{L}_s \rangle$:

$$\Gamma_z = (\mathbf{r} \times \langle \mathbf{F}_{sc} \rangle)_z = r_\perp \frac{ck_0 n_m \alpha''}{\varepsilon_0}\left[\frac{1}{c^2}\langle S_\phi \rangle + (\nabla \times \langle \mathbf{L}_s \rangle)|_\phi\right], \quad (10)$$

with $r_\perp$ the distance to the z-axis. Equation (10) clearly illustrates that the amount of orbital torque transmitted to the particles depends on the radiation pressure and the helicity spatial distribution of the scattered field by means of $\langle S_\phi \rangle$ and $\langle \mathbf{L}_s \rangle$, as well as the optical response of the particle itself by means of $\alpha''$. Interestingly, in the case of the MQ, it is straightforward to show that the contribution of the scattered field to the curl spin force only has an azimuthal component, i.e. it only induces orbital motion. This result is general to any magnetic (TE) multipole field. The interference with the incident illumination leads, however, to important radial and polar components (see Figure 3).

Currently, very few groups[66,80] have investigated optical fields where the effect of spin curl forces can be visibly appreciated in the dynamics of moving nanoparticles in a fluid. In contrast, our calculations directly prove that both spin and radiation pressure contribute to the induced optical torque in the scattered near field of the dielectric cube.

In Figure 3, we show the optical torque (Figure 3a) experienced in the near field by an arbitrary absorbing 40 nm radius spherical nanoparticle with $n \approx 2i$ calculated with Eq.(10) and averaged over several circular rings on parallel transverse planes (perpendicular to the incident propagation direction). Remarkably, particles whose centers of mass are located at different heights experience different contributions from the curl-spin ($\Gamma_z^{Spin}$) and radiation pressure ($\Gamma_z^P$) torques, as can be visually appreciated in the force field plots shown in Figure 3b. Interestingly, the direction of the curl-spin torque can be opposite to the helicity of the incident wave, contrarily to radiation pressure.

In our setup, the particles are initially pushed towards the glass substrate by the incident beam intensity, where they experience a combination of radiation pressure and curl spin torques (Figure 3b (II)). It is worth noting that $\Gamma_z^P$ is nonzero at z=0, contrarily to what one might initially expect from Eq.(7), but we once more emphasize that the total Poynting vector entering in Eq.(10) includes an additional interference term between the incident and scattered field yielding a small azimuthal component.

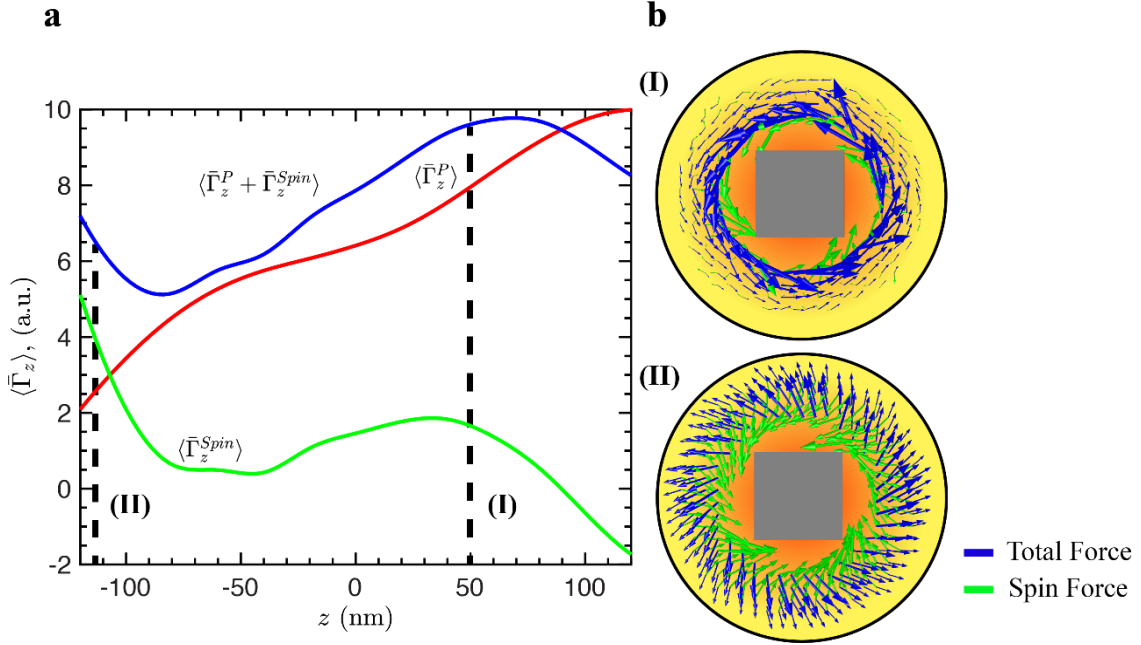

**Figure 3.** (a) Optical torque affecting absorbing 40 nm radius nanoparticles with $n \sim 0.5 + 2i$ in the near field of the MQ resonance ($\lambda = 765$ nm). The time-averaged spin ($\Gamma_z^{Spin}$) and radiation pressure ($\Gamma_z^P$) contributions have been spatially averaged ($\langle \; \rangle$) in parallel x-y planes in the near field along the height of the cube (z axis). (b) Transverse cuts at z=50 nm (I) and z=-120 nm (II) showing the vector field distributions of the x and y components of the total and curl-spin forces around the cube.

To summarize, we have introduced analytical expressions for the optical forces and torque induced on small dipolar absorbing particles, which allowed us to unambiguously distinguish the contributions of the scattering and the spin forces. The numerical calculations presented in Figure 3 demonstrate that both effects mediate the strongly confined (subwavelength) particle rotation with respect to the z-axis (i.e. the direction of propagation of the incident wave).

**4. Nanoparticle dynamics in the optical nanovortex**

We now turn our attention towards the potential applicability of the considered effect as a mixing method for microfluidic reactors. In order to illustrate the concept, the high-index cube is placed in a water host medium ($n_m = 1.335$), containing chemically inert, biologically compatible nanoparticles. The dynamics of the latter will be affected by the optical forces arising due to the interaction with the cube's scattered field together with the Brownian and viscous drag forces induced in the fluid. The obvious and most convenient candidates to act as mixing mediators are gold (Au) nanoparticles, because they would not interact with the chemical and/or biological compounds dissolved in the solutions and are utilized in a broad range of microfluidics applications[81,82]

In order to increase the mechanical orbital torque transferred to the Au nanoparticles and to prevent them from sticking to the walls of the nanocube due to attractive gradient forces, the ratio $\langle \mathbf{F}_{sc} \rangle / \langle \mathbf{F}_o \rangle$ should be maximized. For high enough ratios, non-conservative scattering forces govern the nanoparticle dynamics, causing them to undergo spiral paths around the nanocube and act as stirrers, enhancing convective fluid motion and thus diffusive mixing of any admixtures present in the water solution.

Under the influence of a given optical field, the scattering force can be the leading force acting upon the nanoparticle only if the real part of the nanoparticle polarizability is negligible in contrast to the imaginary one (see Eq.(9)). For simplicity, we assume a spherical shape so that their dipole polarizability can be evaluated analytically with the exact Mie theory formulae by the method described in Refs.[83–85]:

$$\alpha(k_d, R_p) = i\frac{6\pi\varepsilon_0\varepsilon_d}{k_d^3} a_1(k_d R_p), \tag{11}$$

where $k_d$ and $\varepsilon_d$ are the wavenumber and relative permittivity of water, respectively. $a_1$ denotes the first order electric Mie coefficient [86], which depends on the refractive index contrast between the particle and the medium and the dimensionless parameter $k_d R_p$, where $R_p$ is the nanoparticle radius. Figure 4 shows the real and imaginary parts of the polarizability for Au particles of different sizes dispersed in water. The calculations are performed for the well-known optical dispersion properties of bulk Au [87] ($\varepsilon_b^{Au}$), taking into account the Drude size corrections due to the limitation of the electron mean free path in small metallic particles [88]

$$\varepsilon_p^{Au}(\omega) = \varepsilon_b^{Au}(\omega) + \frac{\omega_{pl}^2}{\omega(\omega+i\gamma_b)} + \frac{\omega_{pl}^2}{\omega(\omega+i\gamma_b')},$$
$$\gamma_b' = \gamma_b + \frac{0.7 v_F}{R_p} \tag{12}$$

where $\omega_{pl}, \gamma_b$ and $v_F$ are the plasma resonant frequency, the damping constant from the free electron Drude model, and the Fermi velocity, respectively.

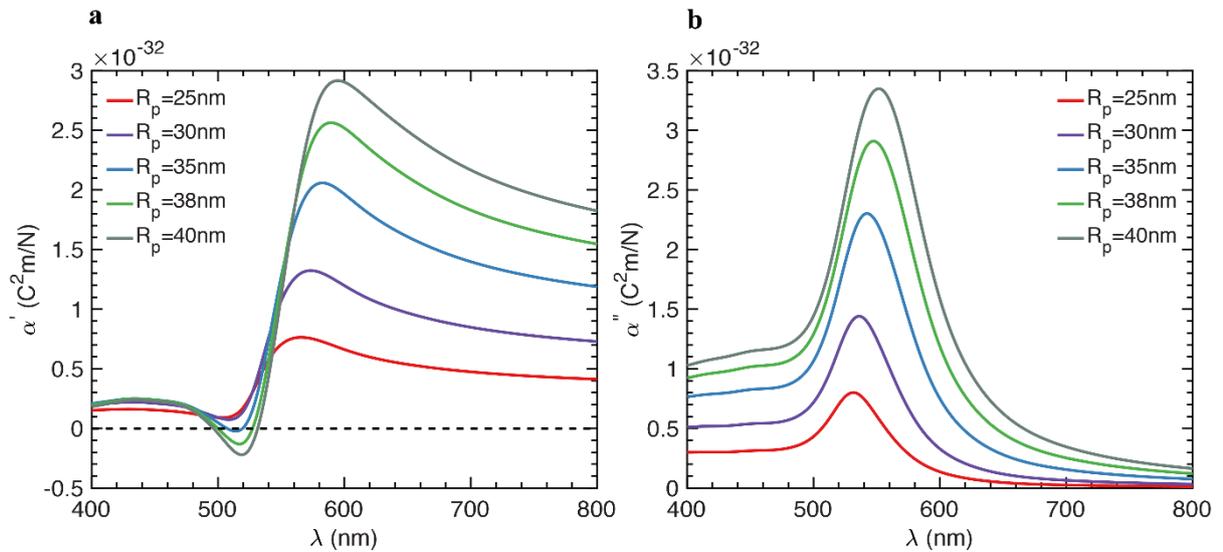

**Figure 4.** Real and imaginary parts of the dipole polarizability are calculated from Eqs. (11) and (12) for Au nanoparticles of different sizes that are dispersed in water. The excitation wavelengths correspond to those in free space. For nanospheres with radius $R_p \geq 35$ nm, the real part can be equal to zero while the imaginary part is enhanced.

Figure 4a demonstrates that in the vicinity of the plasmon resonance nanoparticles with $R_p \geq 35$ nm can fulfill the condition $\alpha' = 0$ with enhanced values of $\alpha''$. For example, for particles with $R_p = 40$ nm, the full suppression of the gradient force occurs at 500 nm and 530 nm (see Figure 4a). Consequently, only scattering forces are allowed for them and the ratio $\langle \mathbf{F}_{sc} \rangle / \langle \mathbf{F}_o \rangle$ is maximized.

Large Au nanoparticles with $R_p \geq 35$ nm could be considered to break the limits of the electric dipole approximation assumed in Eq.(9). To prove its validity for quantitative calculations of the optical forces, we have compared our results with exact numerical computations via integrating the Maxwell stress tensor over a 40 nm radius Au nanoparticle and obtained very good agreement (see Supplementary information). Moreover, the electric field distribution in the system plotted in Figure S1 (B-D) shows negligible perturbations in the presence of an Au nanoparticle with no backscattering, further confirming the validity of the involved approximations.

In order to determine the trajectories of the Au nanoparticles in water, let us consider the radiation pressure coming from the incident beam along the z axis to be completely compensated in the presence of the glass substrate. Therefore, the trajectories can be treated as two-dimensional, localized only in the transverse $x-y$ plane.

Considering scattering force $\langle \mathbf{F}_{sc} \rangle$, viscous drag force $\mathbf{F}_D$ and stochastic Brownian (thermally activated) forces $\mathbf{F}_B$ acting on the nanoparticle of mass $m_p$, the equation of motion can be written in the following form:

$$\langle \mathbf{F}_{sc} \rangle + \mathbf{F}_B + \mathbf{F}_D = m_p \ddot{\mathbf{r}}_p , \qquad (13)$$

where $\ddot{\mathbf{r}}_p$ is the particle instantaneous acceleration vector. Once the scattering force distribution is determined, Eq. (13) can be solved in Comsol Multiphysics© software package utilizing its particle tracing functionality. For small spherical geometries, the Brownian and viscous forces can be expressed as [89]

$$\mathbf{F}_B = \mathbf{\Psi} \sqrt{\frac{12\pi\mu k_B T}{\Delta t} R_p} , \qquad (14)$$

$$\mathbf{F}_D = -6\pi\mu R_p \dot{\mathbf{r}}_p , \qquad (15)$$

where $\mu$ is the dynamic viscosity of water ($8.9 \cdot 10^{-4}$ Pa·s at ambient temperature), $\mathbf{\Psi}$ is a dimensionless vector function of randomly distributed numbers with zero mean [89], $T$ is the temperature of the system and $k_B$ is Boltzmann's constant. The numerical solver models the Brownian forces as a white noise random process with a fixed spectral intensity implying the amplitudes of the force to depend on the iterative time step $\Delta t$ [89]. Formula (15) corresponds to Stokes' law. Its validity is only justified for very low Reynolds numbers [90] being, actually, the case for a microfluidic chip [15]. We assume the system to be at ambient temperature.

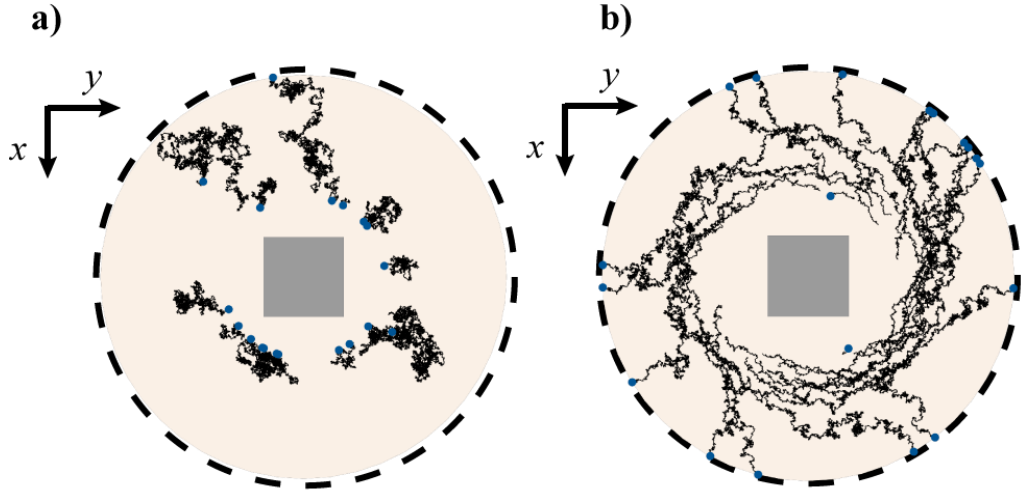

**Figure 5.** Trajectories of Au nanoparticles of 40 nm radius during 1ms of simulations. Illumination wavelength in vacuum was 530 nm wavelength (in water 396 nm). (a) – No incident illumination, only Brownian motion and drag forces act on the particles; (b) – The nanocube is illuminated with a circularly polarized light and the optical force significantly contributes. The Au nanoparticles spirally move around the cube. The figures are scaled to the length of the cube side equal to 158 nm.

The parameters for the simulations are given in Tables S1, S2 of the Supplementary Information. We consider that Au nanoparticles of 40 nm radius are uniformly distributed around the Si nanocube. Their trajectories during a simulation time of 0.1 ms are shown in Figure 5. If the nanocube is not illuminated (Figure 5a), thermal activation induces random movements of the nanoparticles independently on their position in the simulation domain. Conversely, when the system is illuminated with circularly polarized light with intensities in the order of $50-80$ mW/μm$^2$, (corresponding to typical values utilized in conventional optical trapping schemes[91]), a sufficient mechanical torque is transferred to the Au nanoparticles and drives them along spiral trajectories (Figure 5b).

Equation (7) reveals that $\langle \mathbf{S}_\phi \rangle$ becomes negligible far from the nanocube. Furthermore, the numerical simulations show that the curl spin force has no longer a significant effect. Consequently, Brownian motion and conventional radiation pressure dominate the dynamics (see Figure 6b). It is thus natural to introduce the effective radius $r_m$ which specifies the "area of influence" of the optical nanovortex (red dashed curves in Figure 6). For $r \leq r_m$, the majority of the Au nanoparticles circulate around the nanocube, and the dielectric nanocube acts as an effective optical drive for convective stirring of the fluid around it.

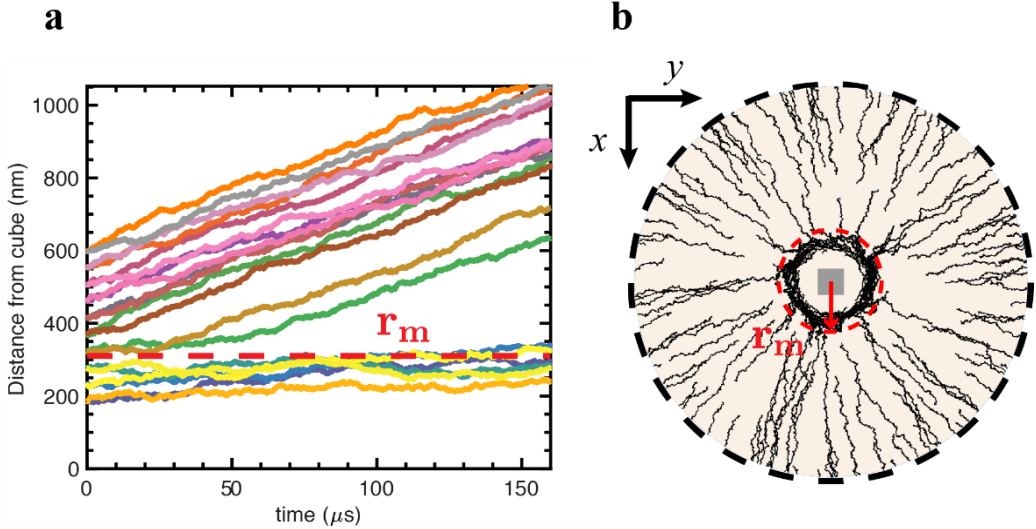

**Figure 6.** (a) Distance from the dielectric nanocube as a function of time for 19 Au nanoparticles initially evenly distributed along the x axis in the simulation domain depicted in Figure 5b. The particle trajectories are calculated up to a simulation time of 0.16 ms. Red dashed curves specify the effective radius $r_m$, where the contribution of optical forces becomes negligible. The optical nanomixing effect is thus achievable at distances from the nanocube smaller than $r_m$.

The radius $r_m$ reaches approximately half of the incident wavelength in water and thus the mechanical effect of optical vortices upon a nanoparticle is formed in the subwavelength region and gets stronger closer to the nanocube. Such a reduced scale cannot be reached using any focused far field, e.g. radial and Bessel beams [63,64,92]. Up to our knowledge, this is the first proposal providing optical nanovortices created in a simple, realizable setup avoiding the need of lossy plasmonic nanoantennas [93,94], short wavelength guided modes [95] or complex chiral structures [96]. Such optical nanovortices represent a promising component for on-a-chip OAM exchange driving light-matter interactions (e.g. controlled light emission from quantum dots [96], super-resolution[97,98] and nanoobject manipulation[63,92]).

## 5. Nanovortex-mediated liquid mixing

To study in detail the liquid flow driven by the proposed nanomixing design, we once again utilize direct time-domain simulation in COMSOL Multiphysics©. At each time step, the particle position and velocity, as well as the fluid pressure and velocity fields are obtained by solving Eq. (13), the Navier-Stokes, and mass balance equations for the fluid [90]. We consider simplified forms of the last two equations assuming laminar, incompressible flow, in accordance with the previous results for the particle trajectories. Furthermore, we impose open boundary conditions at the edges and simulate a large fluid domain around the nanocube (usually lab-on-a-chip microchambers are of the order of tens of micrometers). An Adapted Lagrange-Euler method (ALE)[99] is implemented in order to accurately interpolate the mesh displacements induced by the nanoparticle movement.

Figure 7 shows the calculated stresses and velocity fields in the fluid during $200\,\mu s$. The particles start with zero initial speed and gradually accelerate under the influence of radiation pressure and spin forces arising from their interaction with the optical nanovortex. Consequently, the fluid environment is also displaced, as Figure 7a demonstrates. At longer times, a single-vortex-like velocity distribution is established as shown in Figure 7c.

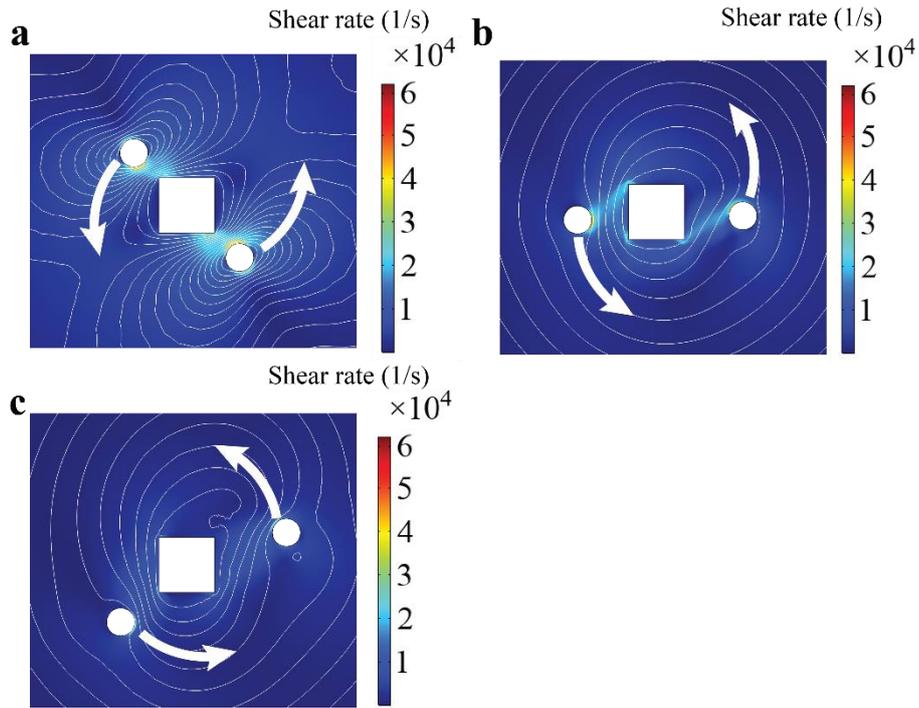

**Figure 7**. Formation of a fluid nanovortex due to the movement of two Au nanoparticles (white circles moving anticlockwise) driven by optical nanovortex formed around the nanocube (white square). Background color map denotes the distribution of stress in the fluid in 1/s and white contours show the fluid velocity field streamlines of the initially static fluid at different times since the nanoparticles became optically driven. (a) t=0.01 $\mu$s, (b) t=100 $\mu$s and (c) t=200 $\mu$s. The represented domain has dimensions $800 \times 800$ nm. Parameters of the simulations are given in Table S1 of the Supplementary information. Open fluid boundary conditions were imposed at a distance of 1.5 $\mu$m from the center of the cube. For the sake of clarity, thermal motion is not taken into account in the simulation.

The velocity streamlines are more inhomogeneous at shorter times, when the nanoparticles start moving. Already at 100 $\mu$s, only small fluid distortions take place very close to the nanoparticles and the nanocube. Therefore, a possible way to further enhance the fluid nanomixing would be to realize periodic switching between left- and right-hand circularly polarized incident light, which would reverse the direction of particle motion maintaining a high level of inhomogeneity in the fluid stress field.

Noteworthy that, while all the previous calculations were performed for Au nanoparticles in the visible range, similar dynamics can also be obtained for Ag nanoparticles in the UV range of the spectrum, where $\alpha' \rightarrow 0$ [100]. Nanomixing in the UV region could be advantageously combined with photochemically active processes of the involved chemical compounds.

## 6. Optical sorting of Au nanoparticles via the nanovortex

Hereinafter, we demonstrate the important capability of the proposed configuration to realize optical force-mediated particle on-chip sorting. In the following paragraphs, we illustrate a novel, dynamical, non-contact size sorting method for Au nanoparticles in liquid solutions addressing one of the most challenging targets of conventional microfluidics with the help of dielectric nanophotonics.

The proposed method is based on the sign reversal displayed by $\alpha'$ close to the plasmon resonance as we demonstrated in Figure 4. The transition reverses the direction of the radial gradient force acting upon the nanoparticle (see the first term in Eq. (9)). At a given incident wavelength, we can split the behavior of the Au nanoparticles into two regions I and II (see Figure 8a). Smaller nanoparticles from region I with positive $\alpha'$ are attracted by the radial gradient force towards the nanocube, while larger nanoparticles, from region II, should be repelled outwards.

However, in region I, there is a competition between the gradient and scattering forces dragging the nanoparticles in opposite radial directions. Simulations performed for Au nanoparticles of radii 20-30 nm proves their movement towards the nanocube (see Figure 8b). For particles smaller than 20 nm, the nanoparticle polarizabilities are very low and thus the driving optical forces are negligible in comparison with Brownian forces. Contrarily, nanoparticles with radii close to or larger than 40 nm (i.e. in the vicinity or inside region II), spiral away from the dielectric nanocube. The particle dimensions in the latter case might go beyond the dipolar approximation expected in Eq. (9). However, our numerical simulations prove the correctness of the previous conclusions (see Supplementary information, Figures S1 and S2).

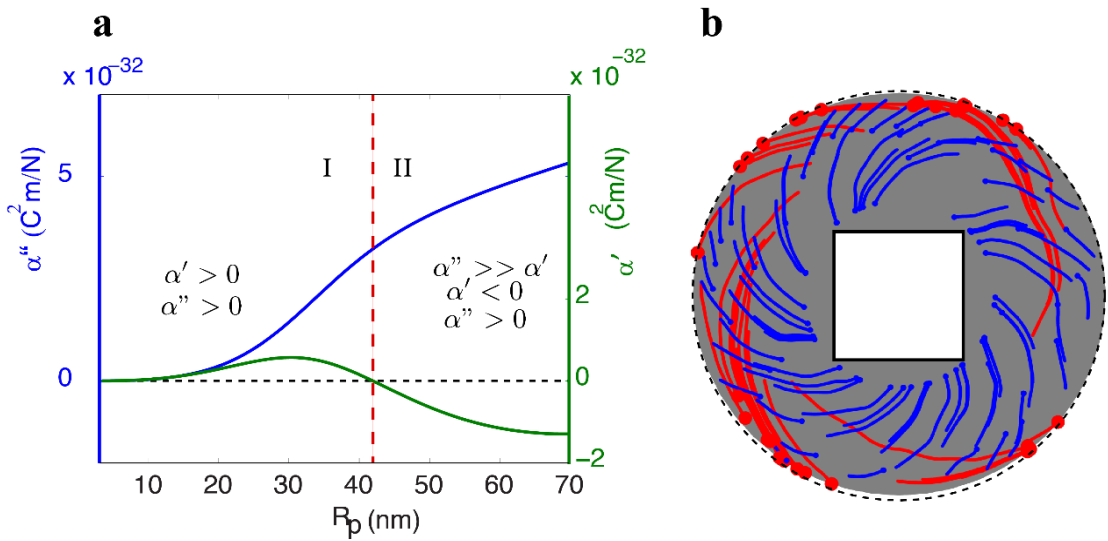

**Figure 8**. (a) Real ($\alpha'$), and imaginary ($\alpha''$) parts of the dipole polarizability as a function of the particle radius $R_p$, for an incident free space wavelength of 530 nm. The black dashed line indicates zero real or imaginary part, while the red dashed line shows the border between region I and II. (b) Calculated trajectories for two sets of particles with radii 20 nm (blue – Region I) and 50 nm (red – region II) demonstrate opposite movement in radial direction.

Figure 8b illustrates the proposed method for nanoparticle separation by comparing the behavior of two sets of Au nanoparticles with $R_p = 20$ nm (blue) and 50 nm (red), respectively. The first set of smaller particles lies well inside region I, while the second one fits the region II. As it can be clearly seen, that gradient forces are strong enough to pull smaller nanoparticles towards the nanocube, tracing an inward curved pattern. Contrarily, strong scattering forces and repulsive gradient ones acting on larger nanoparticles from region II result in an outward motion. Therefore, Au nanoparticles spiral around the nanocube inwards or outwards depending on their size. For the sake of clarity, we have purposely neglected the effect of thermal agitation in Figure 8b, being aware from the simulations that the latter only produces more intricate trajectories without affecting the final outcome of the nanoparticles (not shown here).

A precise, in situ size control of Au nanoparticles is a crucial step in many applications where the processes involved are highly dependent on the latter, e.g. biological cell uptake rates [101,102], toxicity [103] and Raman signal intensity [104].

## 7. Conclusion

We present conditions for maximal conversion of spin angular momentum of the incident light to orbital angular momentum of the scattered light via a specially designed transversely scattering silicon nanocube. The azimuthal component of the Poynting vector of the scattered field originates from the strong magnetic quadrupole resonance excited in the nanocube. A gold nanoparticle of appropriate size, illuminated by such optical field and dispersed in the fluid surrounding the nanocube, experiences a combination of non-conservative spin and radiation pressure forces with non-zero azimuthal component. They are significant only up to a distance of about half of the illuminating wavelength from the nanocube. The exceptionally compact optical nanovortex drives the dynamics of the nanoparticles, inducing a convection fluid flow at the nanoscale. The direction of particle motion can be reversed simply by flipping the helicity of the incident circular polarization illuminating the nanocube. The proposed mechanism can serve as a nanoscale fluid mixer and diffusion booster driven by light in a contact-less and flexible way. Arrays of the studied dielectric structure can be easily imprinted on the surface of a microfluidic chip and controllably illuminated in an independent fashion. Hence, we open the doors to very exciting perspectives such as light controlled mixing or even on-chip directional fluid navigation. Employing the dependence of the optical properties of gold nanoparticles on their size, we demonstrate feasibility to drag nanoparticles affected by the optical vortex either towards or outwards the nanocube. Thus, smaller nanoparticles (20-30 nm in radius) can be aggregated at the nanocube surface while larger nanoparticles move away from it and drive the fluid flow. This behavior can be utilized to perform in situ nanoscale size separation directly inside the microfluidic chip.

The proposed, rather simple concept can be extended to nanostructures of different shapes, to an array of such nanostructures and to different types of dispersed nanoparticles, e.g. silver nanoparticles offer an exciting option to combine optical nanovortex with photo-chemistry at the nanoscale. Our approach opens a new room of opportunities for the integration of simple, optically driven nanosorting or filtering modules in on-chip platforms paving the way towards more efficient functionalities in micro- and nano-fluidic systems.

**Acknowledgements**
ACV, DK, EAG, HKS, DR, SY, and ASS acknowledge financial support from the Russian Foundation for Basic Research (grants 18-02-00414 and 18-52-00005); the force calculations were partially supported by Russian Science Foundation (Grant No. 18-72-10127). PZ acknowledge support of the Technology Agency of the Czech Republic (grant TE01020233) and the Czech Academy of Sciences.

**Conflicts of interest**
The authors declare no conflict of interest.